\documentclass{elsart}
\usepackage{amssymb}
\usepackage{graphicx}

\usepackage[latin1]{inputenc}
\usepackage[british]{babel}
\usepackage[T1]{fontenc}
\newcommand {\simlt}{\lower.5ex\hbox{$\; \buildrel < \over \sim \;$}}
\newcommand {\simgt}{\lower.5ex\hbox{$\; \buildrel > \over \sim \;$}}

\newcommand {\bee}{\begin{equation}}
\newcommand {\ee}{\end{equation}}
\newcommand {\bea}{\begin{eqnarray}}
\newcommand {\eea}{\end{eqnarray}}

\begin{document}
\begin{frontmatter}
\title{From Anisotropy to Omega}

\author{Alessandro Melchiorri\thanksref{1}}
\author{and}
\author{Louise M. Griffiths\thanksref{2}}

\thanks[1]{Dipartimento di Fisica, Universita' degli Studi di Roma 
''La Sapienza'', Ple Aldo Moro 2, Rome, 00185 Italy}
\thanks[2]{Nuclear and Astrophysics Lab, University of Oxford,
Keble Road, Oxford, OX 3RH, UK.}

\begin{abstract}
Following the recent measurement of the acoustic peak 
by the BOOMERanG and MAXIMA experiments in the CMB anisotropy
angular power spectrum, many 
analyses have found that the geometry of the Universe is 
very close to flat, but slightly closed models are favoured.
In this paper we will briefly review how the CMB anisotropies 
depend on the curvature, explaining any assumptions we could make 
and showing that this skewness towards closed models 
can be easily explained by degeneracies in the
cosmological parameters.
While it is difficult to give independent
constraints on the cosmological constant and/or different forms
of dark energies, we will also show that combining CMB
measurements with other observational data will introduce new and tighter 
constraints, like $\Omega_\Lambda > 0$ at high significance.
\end{abstract}
\end{frontmatter}

\section{Introduction}

In the most general and simple inflationary scenario \cite{guth} 
the overall present energy density of the universe must be equal to 
the so-called critical energy density ($\Omega=\rho / \rho_c =1$, 
with $\rho_c=3H_0^2/8 \pi G$). In fact, the
condition for inflation ($\ddot a_{inf} >0$, 
and so $d / dt (H_{inf}^{-1}/a_{inf}) < 0$) is precisely that which 
drives $\Omega(t)$ towards $1$ in the Friedmann equation

\bee
\Omega(t)-1={k \over {a^2(t)H^2(t)}}
\ee

during the inflationary period.
This prediction, however, taken with the {\it standard} CDM 
model of structure formation, is apparently in disagreement with a 
combined set of observations, such as density-velocity galaxy 
field comparisons (\cite{davis}),
pairwise galaxy velocities (\cite{jusk}), X-ray clusters temperature function
evolution (\cite{eke}, \cite{henry}) and velocity dispersions
\footnote{A more conservative approach would say 
that the situation is rather unclear and the X-ray clusters data 
and theoretical modelling can be sufficiently large to prevent an 
unambigous exclusion of $\Omega_m=1$ (see {\it e.g.} 
\cite{cola}, \cite{viana}, \cite{borg}, \cite{blan})}(\cite{carl})], 
which point in favor of a 
low density universe \footnote{From here on, with $\Omega$ we will
indicate only the {\it present} value} ($\Omega_{matter} <1$).
  In order to solve the discrepancy, open ($\Omega=\Omega_{matter} <1$)
inflationary models have been proposed (\cite{sasa}, \cite{buche}
\cite{linde}). This type of model highlights limitations
on the predictiveness of the inflationary scenario, which
is supposed to have the advantage of removing any dependence on 
initial conditions from our present observable universe.
However, even if this picture leads to a more complicated 
phenomenology, it generally determines $\Omega$ directly 
from parameters of the physical theory.

On the other hand, another way to keep low-density models
compatible with the simplest model of inflation is to introduce
a cosmological constant, $\Lambda$, such that  
$\Omega\equiv\Omega_{matter}+\Omega_{\Lambda}=1$.
The presence of such a cosmological constant, which is 
compatible with all the above observations since they are
practically insensitive to it, is also preferred by measurements
of the magnitude-redshift relationship in high-redshift
type Ia supernovae (\cite{perlm}, \cite{riess}).
Nonetheless, this {\it 'natural'} solution
introduces the cosmological constant problem (\cite{weinb1}) 
that is perhaps even more acute in inflationary cosmology 
(see \cite{bran} for a review). 
Thus, an accurate determination of the {\it present} overall 
density parameter $\Omega$, even if not a {\it panacea}
for the cosmological scenario, is at least extremely important in 
understanding which theoretical framework could explain 
the above conflicts.

There is much experimental evidence for
 the presence of a peak in the
CMB angular power spectrum (\cite{mille},
\cite{mausk}).  Furthemore, with the recent release of the 
BOOMERanG-98 \cite{debe} and MAXIMA-1 \cite{hano} spectra, the
shape and position of the peak has been detected with 
unprecedented accuracy. 
This result, apart from being in wonderful agreement with the
standard scenario of primordial {\it adiabatic} fluctuations, 
has important consequences on the parameter $\Omega$ 
(\cite{melk00}, \cite{lange}, \cite{balbi}, \cite{jaffe}, \cite{ruth}). 
As previously noted and already well explained in the literature,
the effect of the curvature is to change the relationship
between the physical scales on the Last Scattering Surface
(LSS) and the corresponding angular scales (see., {\it e.g.} 
\cite{weinb}, \cite{turn}).
In an open universe, for example, the geodetics focalize in such
a way that a particular angular scale will correspond to
a greater physical scale on the LSS than the one expected in a flat model.
The immediate result is a shift in the radiation angular power spectrum
(the so-called $C_\ell$'s),
and thus a dependence of the position of the first peak 
$\ell_{peak}$ on the curvature (\cite{wils},
\cite{kamion}, \cite{silk}, \cite{hu}, \cite{scott}). 
The CMB power spectrum is therefore a powerful tool
for the determination of the curvature and so of the overall
energy density.

\begin{figure}[htb]
\includegraphics[width= 10cm,height=8cm,angle=-90]{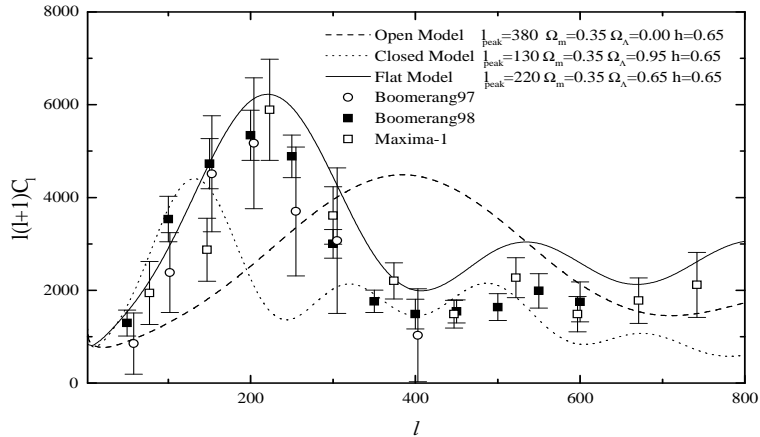}
\caption{The BOOMERanG and MAXIMA $C_{\ell}$'s data together with the most
viable open, closed and flat models from present non--CMB observations.}
\label{figregion}
\end{figure}

\section{$\Omega$ and the shift of the CMB Angular Power Spectrum}

In Fig.1, we show the recent data from the BOOMERanG
(\cite{mausk},\cite{debe}) and from the MAXIMA \cite{hano} 
experiments together with the predictions of the most viable
open, flat and closed adiabatic models (from present non--CMB
observations).
It is quite evident that the open (closed) model predicts a
first peak on smaller (larger) scales with respect to the flat model, 
which is in much better agreement with the data.

In the literature, the dependence of $\ell_{peak}$ 
on $\Omega$ is often expressed as 

\bee
\ell_{peak}\sim{220 \over \sqrt{\Omega}}
\ee

However, this approximation is not correct in $\Lambda$--dominated 
universes (see Fig.1, where the peak for the closed model
is at $\sim 130$ instead of $\sim 190$) and does 
not take into account the further dependence of
$\ell_{peak}$ on other parameters like the Hubble 
constant and the matter density $\Omega_m$.
So, in view of the recent 
$\ell_{peak}=197 \pm 6$ (\cite{debe}), further modifications
to the above formula are needed (see also \cite{wein2}, \cite{hu2}).
Also, the width of the peaks and the inter peak distance vary as 
functions of $\Omega$ (again see Fig.1) so a more complete 
expression is needed to describe these effects.  
This point is rather important because knowing the exact 
dependence of the CMB spectrum on $\Omega$ will help us in understanding
the shape of the probability distribution function for this parameter
and, ultimately, how well it can be measured independently.

In order to do this, it is convenient to introduce  the 'shift' 
parameter $R$ defined as 

\bee
\ell_{peak}=\ell^{flat}_{peak}/{\it R}
\ee

where $flat$ indicates the flat, pure--CDM, $\Lambda=0$ model.
The use of the $R$ parameter is more appropriate than the 
conventional $\ell_{peak}$ because it has a clearer
geometrical dependence.

As usual, let us assume the metric of space--time to be of a 
Friedman--Lemaitre--Robertson--Walker (FLRW) form with curvature $k$:

\bee
ds^2=a(t)^2[-d\eta^2+\gamma_{ij}dx^idx^j]
\ee

with

\bee
\gamma_{ij}dx^idx^j=dr^2+\chi^2(r)(d\theta^2+sin^2\theta\phi^2)
\ee

where the function $\chi(r)$ depends on the curvature $k$ and
is $r$, $\sin(r)$ or $\sinh(r)$ for $k=0$ (flat models), 
$k=1$ (closed models) and $k=-1$ (open models), respectively.
The position of the first acoustic peak is determined primarly
by the angle subtended by the acoustic horizon $\lambda_{ac}$ at
decoupling time, $\eta_{dec}$.
The angle under which a given comoving scale $\lambda$ at
conformal time $\eta_{dec}$ is seen on the sky is given by
$\theta(\lambda)=\lambda/\chi(\eta_0-\eta_{dec})$.
As the harmonic number $\ell$ is inversely proportional
to the angle $\theta$, we have  $R=\theta_{ac}/\theta_{ac}^{flat}$, 
with $\theta_{ac}=c_s\eta_{dec}/\chi(y)$, where 
$c_s=1/\sqrt{3(1+3\Omega_b/4\Omega_{rad}(1+z_{dec}))}$
denotes the adiabatic sound speed of the baryon/photon
plasma at decoupling.
It is possible to show that \cite{turn}:

\bee
\eta_{dec}={{2 \sqrt{|\Omega_k|}} \over {\Omega_m}} (\sqrt{\Omega_{rad}+
\Omega_m/(z_{dec}+1)}-\sqrt{\Omega_{rad}}) 
\ee 

where $z_{dec}\sim1100$ is the redshift at decoupling and $\Omega_k=
1-\Omega_m-\Omega_{\Lambda}$.
Furthermore, we have:

\bee
y=\eta_0-\eta_{dec}=\sqrt{|\Omega_k|}\int_0^{z_{dec}} { {dz} \over
{[\Omega_m(1+z)^3+\Omega_k(1+z)^2+\Omega_{\Lambda}]^{1/2}}}
\ee

For a flat, $\Omega_m=1$, $\Omega_{\Lambda}=0$, universe we have

\bee
\theta_{ac}^{flat} =c_s\eta_{dec}/(\eta_0-\eta_{dec})=
c_s(\sqrt{\Omega_{rad}+1/(z_{dec}+1)}-\sqrt{\Omega_{rad}})
\ee

We then find, keeping constant $\Omega_bh^2$ (and so $c_s$) 
and $\Omega_mh^2$ (see next section) \cite{efst}:

\bee
R={{2 \over \chi(y)}}\sqrt{{|\Omega_k| \over \Omega_m}}
\ee

which is a quantity just dependent on 
$\Omega_k=1-\Omega_\Lambda-\Omega_m$ and $\Omega_m$.

\begin{figure}[htb]
\includegraphics[width= 10cm,height=8cm,angle=-90]{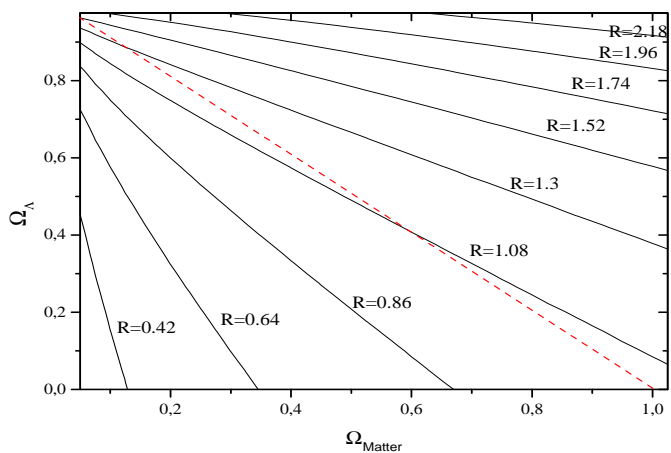}
\caption{$R=constant$ lines in the $\Omega_\Lambda-\Omega_m$ plane.}
\label{figregion}
\end{figure}

In Fig.2 we plot the contours at $R=const$ in the
 $\Omega_{\Lambda}-\Omega_m$ plane, together with 
one dashed line such that $\Omega=1$.
As we can see, open models have in general $R < 1$, 
while closed models have $R > 1$. This is the
usual result that in an open model the peaks are
shifted towards greater $\ell$ values (smaller angular scales) with respect
to the flat model case, while for closed models we have the 
opposite effect. It is worth noting that the CMB angular shift is
not linearly related to the curvature: 
lines at $\Omega=const$ are not parallel with the $R=const$
contours, but have multiple intersections, especially in 
regions far from flatness (again, see Fig.2).

In Section 4, we will build a
likelihood distribution function $L$ for $\Omega$, 
using a Bayesian approach to compare the current CMB data
with the theoretical predictions and then marginalizing over the
remaining cosmological parameters.
The above result implies that the 
probability distribution function for $\Omega$
(if we assume a flat prior distribution on the remaining 
parameters while marginalizing) will always be 'skewed' so 
that $\Omega$ will never be measured at a level 
better than $10-20 \%$\footnote{Another interesting point, is that for $\Omega_m \sim 0$
the lines at $R$=constant converge towards $\Omega_\Lambda=1$, but
we must warn the reader that in the de Sitter 
solution the notion of open, flat or closed Universe becomes
ambigous \cite{dolg}.}.

\section{The geometrical degeneracy}

With the parameter $R$ fixed, 
the structure and position of the $C_\ell$ spectrum is 
dependent on $2$ physical scales: the equality scale and the sound
horizon at decoupling scale.
These quantities are completely defined once we choose the
abundance of cold dark and baryonic matter in our model,
by the parameters: $\omega_{cdm}=\Omega_{cdm}h^2$ and
$\omega_b=\Omega_bh^2$.
The CMB spectrum also depends on
the characteristics of the primordial inflationary
perturbations. Assuming that we have already selected
the primordial power spectrum of our model, both in shape 
(tilted or 'blue') and in nature (adiabatic,
isocurvature, hybrid), the structure of the CMB angular
peaks is completely determined by $R$, $\omega_b$ and
$\omega_m=\omega_b+\omega_{cdm}$.
cdm
This result has an important consequence: if we let $\Omega_b$,
$\Omega_{cdm}$ and $h$ assume any value but a fixed 
$\omega_m$ and $\omega_b$, the lines at $R=const$ in the 
$\Omega_m-\Omega_{\Lambda}$ 
plane correspond to sets of degenerate power spectra 
with an {\it identical} shape on subdegree 
angular scales ($\ell > 30$) 
\cite{efst}.

\begin{figure}[htb]
\includegraphics[width= 10cm,height=8cm,angle=-90]{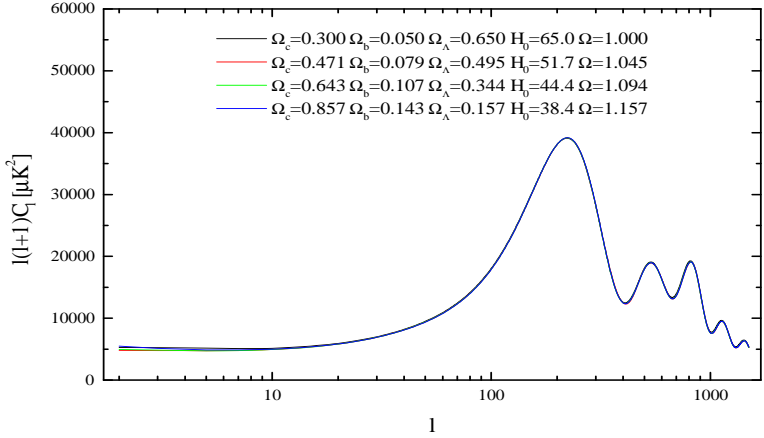}
\caption{Concordance model and degenerate models}
\end{figure}

In Fig.3 we draw a set of degenerate models with $R$, $\omega_m$ and
$\omega_b$ set to those of the cosmological
concordance model, $\Omega_{m}=0.35$, $\Omega_{\Lambda}=0.65$,
$h=0.65$, $\omega_b=0.021$, $n_s=1$.
It is clear from the degeneracy of the models 
that, once $\Lambda$ is included,
the peak position is not directly related to $\Omega$.
It is also clear that it seems impossible to obtain any relevant 
and independent information from the CMB on $\Omega_{\Lambda}$:
the Integrated Sachs-Wolfe effect on large scales could  in 
principle break the degeneracy but cosmic variance 
and a possible presence of a gravity wave
background make this effect difficult to disentangle.

Our main results are then the following:

\begin{itemize}
\item Lines at $R=const$ in the $\Omega_{\Lambda}-\Omega_m$ plane
correspond to sets of degenerate $C_\ell$ power spectra.
\item Given a flat $\Omega_{\Lambda}\sim 0.65$
 model, a degenerate {\bf closed} model can be found by decreasing 
$\Omega_\Lambda$ and $h$ and increasing $\Omega_b$ and $\Omega_{cdm}$.
\item $R$, $\omega_m$ and $\omega_b$ are the most meaningful
 CMB anisotropy observables.
\item The CMB spectrum is a useful tool for the determination
of $\Omega$ only if we live in a flat universe ($R=const$ are 
not parallel to $\Omega=const$ when $\Omega$ greatly 
differs from one).
\item Assuming that the concordance model describes our real 
universe we expect that a likelihood analysis for $\Omega$, 
using only the CMB power spectrum and without including any 
external information about $h$, $\Omega_m$ and $\Omega_\Lambda$,
will always be skewed towards closed models.
\end{itemize}

\section{From CMB to $\Omega$}

Let us now describe the standard tools for extracting
the cosmological parameters from CMB anisotropy
observations. Here we will analyze the recent 
BOOMERanG97 (\cite{mausk}), BOOMERanG98 (\cite{debe}) 
and MAXIMA (\cite{hano}) results.
The power spectra from these experiments were estimated in 
$7$, $12$ and $10$ bins respectively, spanning the range
$25 \le \ell \le 785$. In each bin, the spectrum is assigned
a flat shape, $\ell(\ell+1)C_{\ell}/2\pi=C_B$.
Following \cite{bjk00} we use the offset lognormal approximation
to the likelihood $L$. In particular we define:

\begin{equation}
-2{\rm ln} L
=(D_B^{th}-D_B^{ex})M_{BB'}(D_{B'}^{th}-D_{B'}^{ex}),
\end{equation}
\begin{equation}
D_{B}^{X}={\rm ln}(C_{B}^X+x_B),
\end{equation}
\begin{equation}
M_{BB'}=(C_B^{ex}+x_B)F_{BB'}(C_{B'}^{ex}+x_{B'}),
\end{equation}

where $C_B^{th}$ ($C_B^{ex}$) is the theoretical (experimental)
band power, $x_B$ is the offset correction and $F_{BB'}$ is
the Gaussian curvature of the likelihood matrix at the peak. 
Of course, $C_B^{th}$ will depend on the various parameters 
of our cosmological model ($\omega_m$,$n_s$, ...),
 and so it will be the likelihood function $L$.
In order to compute the likelihood for a given parameter only
$\alpha$ we can either {\it marginalize} over all the remaining
parameters, namely carry out the integral
\bee
L_{marg}(\alpha)=\int P_{prior}(\alpha,{\vec \Pi})L(\alpha,{\vec \Pi})d{\vec \Pi}
\ee
where $\vec \Pi$ is a vector containing all the remaining parameters
and $P_{prior}$ is the prior probability distribution, or
we can {\it maximize} {\it i.e.} for a fixed $\alpha$ find the 
${\vec \Pi_{max}}$ wich maximizes 
\bee
L_{max}(\alpha)=P_{prior}(\alpha, {\vec \Pi_{max}}) L(\alpha,{\vec \Pi_{max}}).
\ee
The two methods, in general, agree at a level of $\sim 10 \%$.
The maximization method is usually based 
on a search algorithm through the second derivative
of the likelihood matrix (\cite{d&k00}). In this
approach the $C_{\ell}$ spectrum is computed on the way,
without sampling the whole parameter space. 
A different approach is based on building a database of 
$C_{\ell}$'s on a discretized grid of the parameter space. 
$L(\Omega)$ is then obtained by maximizing and/or 
integrating the likelihood computed on the grid 
(\cite{melk00}, \cite{tegmark00}). 
Of course, producing a grid of models can be quite 
computationally expensive even with the new and fast boltzmann 
codes like CAMB \cite{CAMB} or CMBFAST \cite{CMBFAST}. 
But this problem can be drastically 
reduced using morphing \cite{sigurdson} or interpolation 
\cite{tegmark01} algorithms.

The definition of the database is rather important because it
defines the {\it internal} $P_{prior}$ of our analysis and, in
general, it is better to have this prior as flat
as possible for each given parameter. 
This brings us to the choice of the variables in which
the database must be sampled. As we saw in the previous section,
the CMB anisotropies are mainly sensitive to the {\it physical}
variables $\omega_i$ and $R$, so sampling in those variables
will avoid degeneracies. Furthemore, the physical baryon
density $\omega_b=\Omega_bh^2$ is well determined by 
independent measurements like primordial nuclide
abundances, so this is the optimal choice for 
extracting information about this parameter (without involving
complicated Jacobian transformations) or assume external priors 
for it. For the same reason, extracting confidence limits
on parameters like $h$ can be a little more elaborate 
with this sampling, being the database in $\sim h^2$.

Another possibility is to sample the database in 
{\it cosmological} variables, like $\Omega_i$ and $h$.
Of course, this will introduce degenerate models in our database
but this sampling has the advantage of obtaining direct
constraints on the commonly used parameters and with flat prior 
distributions.
In most of the recent papers, either a 'hybrid' variables approach,
with sampling in $\omega_i$ and $1-\Omega_m-\Omega_{\Lambda}$, or 
the {\it cosmological} variables approach has been used.
Here we will choose the database approach, sampling the
parameter space in {\it cosmological } variables as follows:
$\Omega_{m}=0.1, ...,1.1$; $\Omega_{\Lambda}=0.0, ..., 1.0$; 
$\Omega_{b}=0.01, ..., 0.25$;
$h=0.25, ..., 0.95$ and
$n_s=0.50, ..., 1.50$. We will not consider the possibility of 
high redshift reionization of the intergalactic medium $\tau_c >> 0$, 
a gravity waves contribution or the effect of massive neutrinos.

\section{Removing the geometrical degeneracy: Results}

\begin{figure}[htb]
\centering
\includegraphics[width= 7cm,height=9cm]{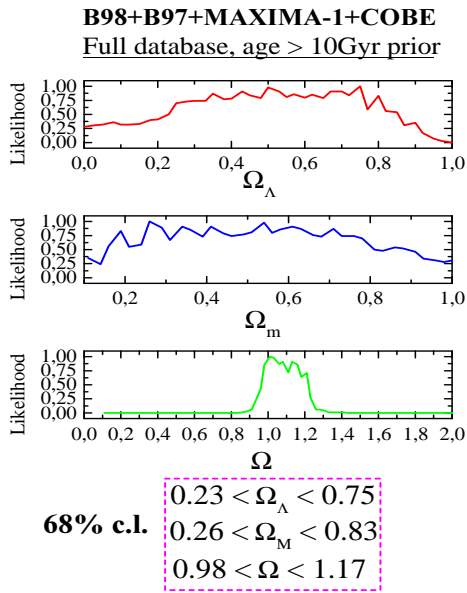}
\caption{Likelihood functions with age prior.}
\end{figure}

\begin{figure}[htb]
\centering
\includegraphics[width= 7cm,height=9cm]{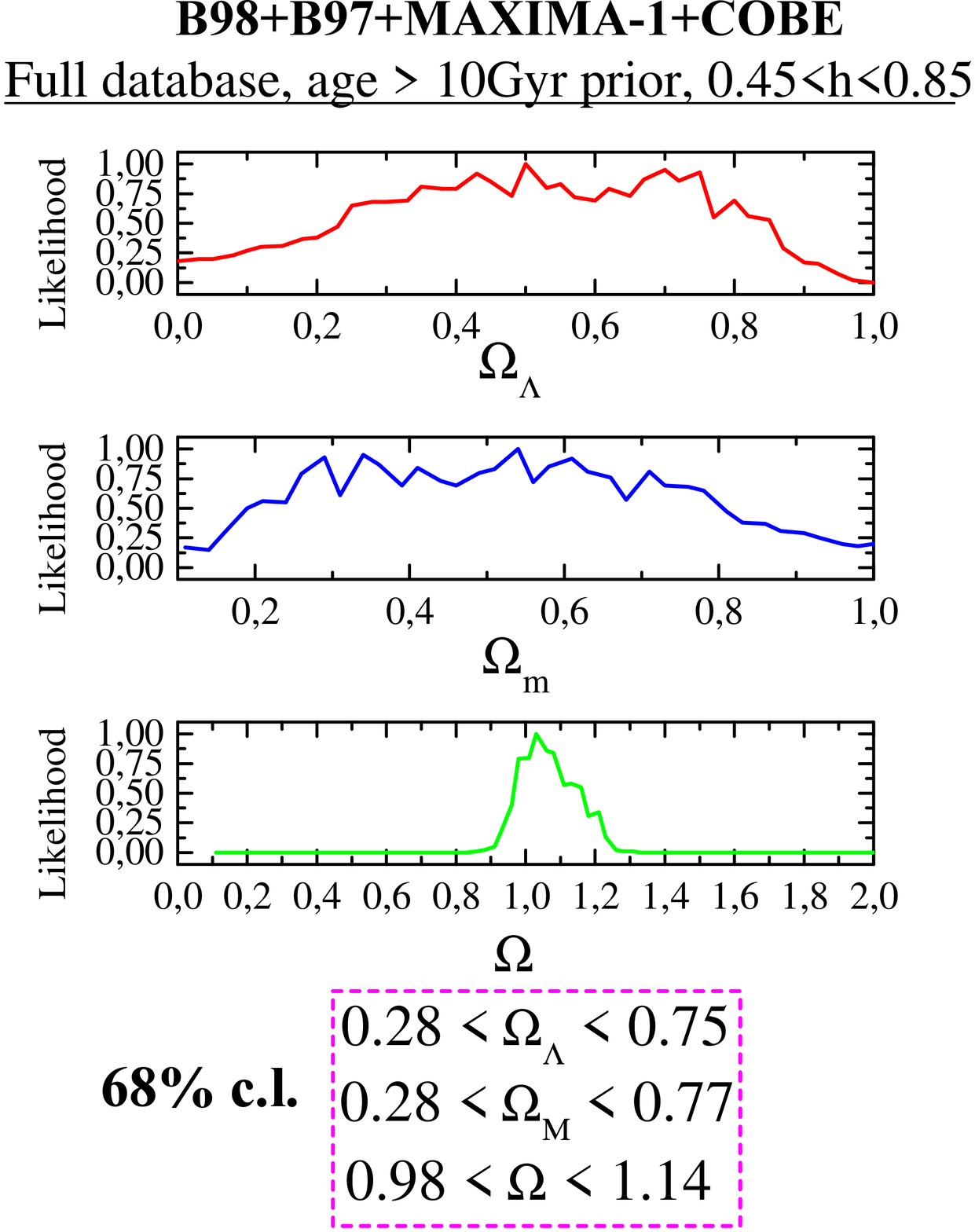}
\caption{Likelihood functions with $h=0.65\pm 0.2$ and age priors.}
\end{figure}

\begin{figure}[htb]
\centering
\includegraphics[width= 7cm,height=9cm]{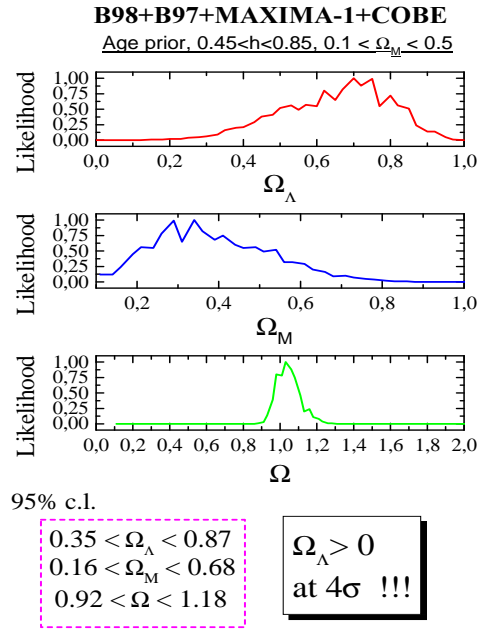}
\caption{Likelihood functions with $\Omega_m=0.3 \pm 0.2$ prior.}
\end{figure}

\begin{figure}[htb]
\centering
\includegraphics[width= 7cm,height=9cm]{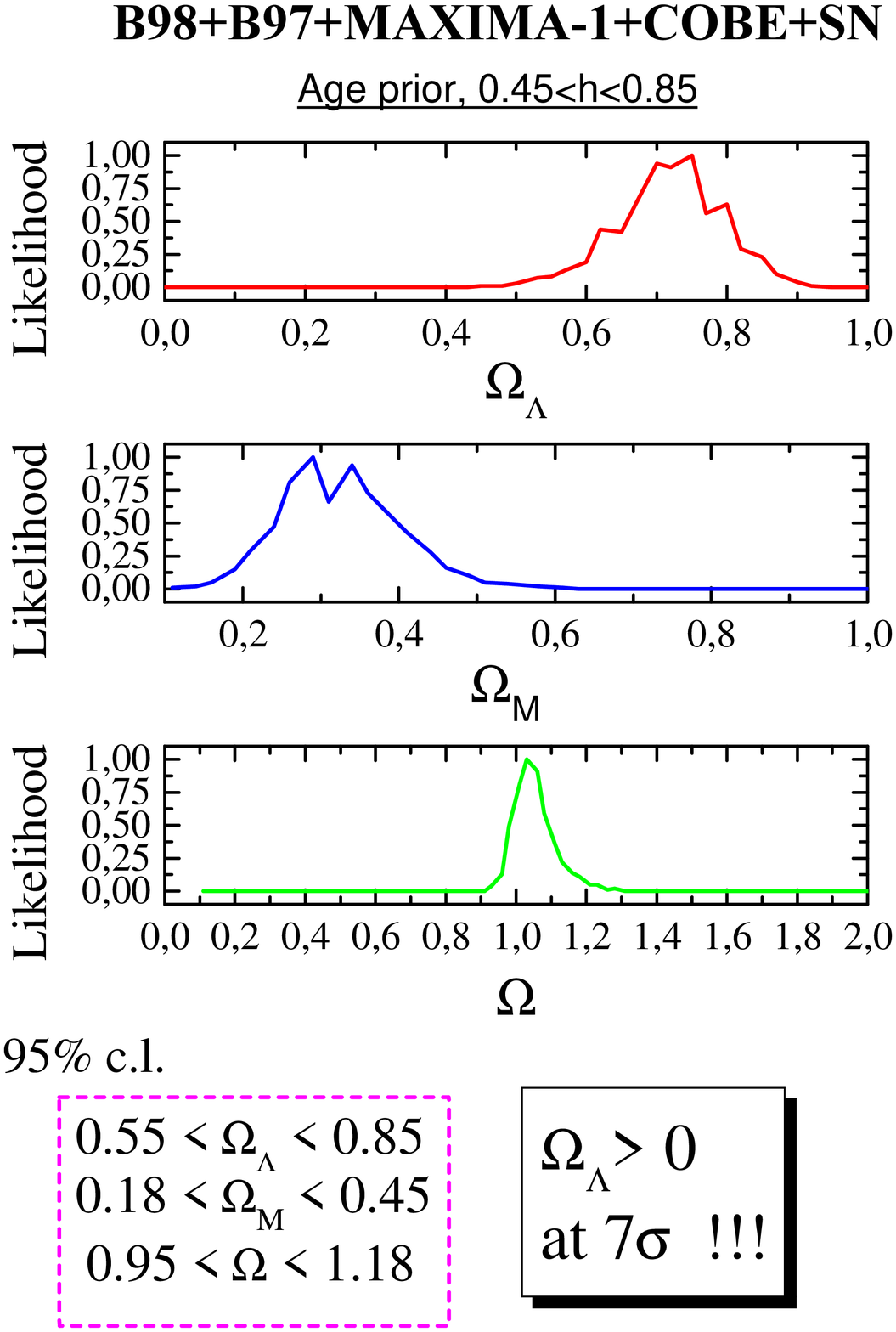}
\caption{Likelihood functions for a combined analysis with Supernovae.}
\end{figure}

In Fig.4 we plot our likelihood contours for $\Omega$, $\Omega_m$ and
$\Omega_{\Lambda}$  using just the intrinsic internal
priors of the database plus the quite reasonable age prior
$t_{universe}> 10 Gyr$. 
As we can see, the likelihood for $\Omega$ is skewed towards closed 
models but is  consistent with flatness. 
It is rather important to note that this skewness is largely due
to the $R$-degeneracy which makes 'more' closed models
compatible with the observations.
The likelihood for $\Omega_{\Lambda}$ and $\Omega_m$ are quite
flat due to the geometrical degeneracy but they nonetheless
feel border effects from the database priors.
The likelihood for $\Omega$ starts to be in even more agreement with
flatness when a Gaussian prior $h=0.65 \pm 0.2$ is assumed as in 
Fig.5.
This clearly shows that most of the degeneracies
in the $h < 0.5$ region are well removed by the prior.
Including a prior $\Omega_m=0.3 \pm 0.2$ (Fig.6) as suggested
by the majority of measurements, shrinks the likelihood towards
$\Omega=1$ and gives a strong determination for the
cosmological constant $\Omega_{\Lambda}>0$ at $4\sigma$ level.
The complementarity with the supernovae type Ia measurements
is even more clear in Fig.7, where a combined 
CMB+SnIA analysis gives $\Omega_{\Lambda}>0$ at more than
$7\sigma$ and $\Omega =1$ with a few percent uncertainty.

\section{Conclusions}

The BOOMERanG and MAXIMA data support the main
prediction of the inflationary paradigm: that the
geometry of the universe is flat.
The small deviations towards closed models reported
in various analyses (\cite{lange},\cite{jaffe})
can be easily explained by the
degeneracies in the cosmological parameters, which
make {\bf more closed} models compatible with
the data. These conclusions are considerably
strengthened by the inclusion of other cosmological
data such as measurements of the Hubble constant,
 the overall matter density $\Omega_m$ and the accelerating 
expansion rate indicated by observations of distant
Supernovae.
At the same time, $\Omega_\Lambda$ and other forms 
of 'dark energy' cannot
be well determined by the CMB data alone, in spite
of their high precision.
This does not mean that CMB measurement are not
useful in the determination of such parameters: 
combining the CMB data with constraints from
observations of large-scale-structure and from observations
of SN-Ia increases the extent to which $\Lambda$ can be
quantified, with $\Omega_{\Lambda}>0$ at $\sim 7 \sigma$.
Future data from the PLANCK and/or SNAP satellites 
will hopefully enable us to resolve the 'dark energy' puzzle.

\section*{Aknowledgements}
\vskip-.1in
We would like to thank Monique Signore, Paolo de Bernardis, 
Marian Douspis, Ruth Durrer, Pedro Ferreira, Graca Rocha, Joe Silk,
Filippo Vernizzi and Nicola Vittorio.
We also acknowledge using CMBFAST, CAMB and Lloyd Knox's 
RADPack \cite{RADPACK} packages for the COBE data.

\end{document}